\documentclass[aps,prl,twocolumn,nofootinbib,superscriptaddress]{revtex4-1}
\usepackage{amsmath}
\usepackage{amsfonts}
\usepackage{amssymb}
\usepackage{dcolumn}
\usepackage{bm}
\usepackage{graphicx}
\usepackage{adjustbox}
\usepackage{epstopdf}
\usepackage{color}
\usepackage{xcolor}
\usepackage{textcomp}
\usepackage{gensymb}
\usepackage{adjustbox}
\usepackage{ulem}
\usepackage{comment}
\definecolor{deep-blue}{rgb}{0.17, 0.17, 0.89}

\newcommand{\smallindice}[1]{_{\mbox{\textit{\tiny{#1}}}}}

\def\BNOO {Ba$_2$NaOsO$_6$}
\def\BCOO {Ba$_2$CaOsO$_6$}
  
 \def\BNCOO {Ba$_2$Na$_{1- \rm{x}}$Ca$_{\rm x}$OsO$_6$}

\def\uSR {$\mu$SR}
\def\ie {{\it i.e.} }

\begin{document}

%\title{Electron doped Mott insulator with strong spin-orbit coupling \BNCOO{}}
\title{Effects of charge doping on Mott insulator with strong spin-orbit coupling, \BNCOO{}}

\author{E. Garcia}
\affiliation{Department of Physics, Brown University, Providence, Rhode Island 02912, USA}
\author{R. Cong}
\affiliation{Department of Physics, Brown University, Providence, Rhode Island 02912, USA}
\author{P. C. Forino}
\affiliation{Department of Physics and Astronomy "A. Righi", University of Bologna, I-40127 Bologna, Italy}
\author{A. Tassetti}
\affiliation{Department of Physics and Astronomy "A. Righi", University of Bologna, I-40127 Bologna, Italy}
\author{G. Allodi}
\affiliation{Dipartimento di Scienze Matematiche, Fisiche e Informatiche, Universit{\'a} di Parma I-43124 Parma, Italy}
\author{A. P. Reyes}
\affiliation{National High Magnetic Field Laboratory, Tallahassee, Florida 32310, USA}
\author{P. M. Tran}
\affiliation{Department of Chemistry and Biochemistry, The Ohio State University, Columbus, Ohio 43210, USA}
\author{P. M. Woodward}
\affiliation{Department of Chemistry and Biochemistry, The Ohio State University, Columbus, Ohio 43210, USA}
\author{C. Franchini}
\affiliation{Department of Physics and Astronomy "A. Righi", University of Bologna, I-40127 Bologna, Italy}
\author{S. Sanna}
\affiliation{Department of Physics and Astronomy "A. Righi", University of Bologna, I-40127 Bologna, Italy}
\author{V. F. Mitrovi\'c}
\affiliation{Department of Physics, Brown University, Providence, Rhode Island 02912, USA}

\date{\today}
\begin{abstract}
 {\bf The effects of   doping on the electronic evolution of the Mott insulating state have been extensively studied in efforts to  understand mechanisms of  emergent quantum phases of materials. %In systems with significant spin-orbit coupling (SOC), the joining of SOC with strong electron correlations and charge doping can induce   exotic phases. 
 The study of these effects becomes ever more intriguing in the presence of entanglement between  spin and orbital degrees of freedom. 
 Here, we present a comprehensive investigation of  charge doping  in the double perovskite \BNOO,    a complex Mott insulator where such entanglement plays an important role.   We establish that the  insulating magnetic ground state evolves from canted antiferromagnet (cAF) \cite{Lu_NatureComm_2017} to N{\'e}el order for dopant levels  exceeding $\approx 10 \%$. %, indicating the enhancement of electron-electron correlations with doping. 
 Furthermore, we determine that a broken local point symmetry (BLPS) phase,   precursor to the  magnetically ordered state \cite{Lu_NatureComm_2017}, occupies  an extended  portion of the  ($H{-}T$) phase diagram with increased doping. This finding reveals that the breaking of the local cubic symmetry is driven by a multipolar order, most-likely of the 
antiferro-quadrupolar type \cite{khaliullin2021exchange,churchill2021}.}
%
% 
%ferro-octupolar type. }

 \end{abstract}

\maketitle

%%%%%%%%%%%%%%%%%%%%%%%%%%%%%%%%%

%\section{Introduction}
%\label{sec-introduction}
%  \vspace*{-0.20cm}

 %%%%%%%%%%%%%%%%%%%%%%%%%%%%%%%%%%%%%%%%%%%%%%%%%%%%%%%%%%%%%%%%%%%
%%%%%%%%%%%%%%%%%%%% INTRODUCTION %%%%%%%%%%%%%%%%%%%%%%%%%%%%%%%%
%%%%%%%%%%%%%%%%%%%%%%%%%%%%%%%%%%%%%%%%%%%%%%%%%%%%%%%%%%%%%%%%%%%
\noindent {\bf Introduction}

\noindent Intricate interplay  between strong   electron correlations, and intertwined spin and orbital degrees of freedom leads to many diverse complex quantum phases of matter %, such as multipolar order 
\cite{witczak2014correlated, jackeli09mott,PhysRevResearch.3.043016,doi:10.7566/JPSJ.90.062001}. Often, correlations of the spin and orbital degrees of freedom can be treated on distinct energy scales. However, this is not the case
in systems   containing $5d$ transition-metal ions, where spin-orbit coupling (SOC) and electron correlations are   comparable in size \cite{Chen_PRB_2010, Chen:2011,svoboda2021orbital,jackeli09mott, MonteCarlo_2014magnetism, cong2021monte, PhysRevX.11.011013, PhysRevResearch.4.013062}. As a result, $5d$ compounds exhibit a wide range of exotic magnetic properties, structural distortions, and multipolar ordering \cite{witczak2014correlated,Lu_NatureComm_2017,PhysRevB.97.054431,PhysRevMaterials.5.104410,paramekanti2020octupolar,voleti2020multipolar,churchill2021,PhysRevResearch.2.022063,PhysRevB.100.214113,maharaj2020octupolar}.  The underlying physical ground state, is controlled by the multiplet structure of the constituent ions,  the nature of the chemical bonds in the crystal, and its symmetry. This complexity often leads to intricate  quantum ``hidden'' orders, elusive to most standard experimental probes. 
Nevertheless, the structural, magnetic, and electronic properties can be finely tuned by altering the  degeneracy of a multitude of ground states   varying  external perturbations, such as pressure, strain and doping \cite{PhysRevB.100.214113,voleti2021octupolar,kesavan2020doping,2022polaron}.

The expectation, under the simplest picture, is that Mott insulators with integer number of electrons per site  favor an antiferromagnetic ground state (AFM), and that charge doping %the system 
 leads to a metal-insulator transition (MIT)  into a conducting state \cite{RevModPhys.70.1039}. 
The most notable example of   doping   is the superconducting state in cuprates believed to emerge  from  a parent antiferromagnetic   Mott state \cite{lee2006doping,SCMI_Kivelson21,Kivelson22}. Specifically,  as  doping  increases, antiferromagnetism gives way to exotic orders such as ``stripe'', unidirectional charge density wave, spin density wave, and unconventional $d$-wave superconductivity   and  with high enough doping the system becomes a Fermi liquid \cite{lee2006doping,RevModPhys.87.457, Kivelson22}. In addition to this well known class of MITs induced in Mott insulators  by Coulomb interactions \cite{middey2016physics,tokura2000orbital, xia2020research,lee2006doping,SCMI_Kivelson21},   insulators purely driven by spin correlations have been recently observed \cite{MIT_SpinCorr}.
Yet another interesting case arises in Mott insulators when a strong SOC locally entangles  the spin and orbital degrees of freedom. In such systems unconventional quantum magnetic and multipolar orders may stabilize \cite{jackeli09mott,PhysRevResearch.3.043016,paramekanti2020octupolar,voleti2020multipolar,voleti2021octupolar}. 
Furthermore, the effects of charge doping are expected to be strikingly different than in systems    where SOC can be treated as a perturbation to electronic correlations \cite{jackeli09mott, carter2013theory, han2016enhanced,bandyopadhyay2019evolution, vogl2018interplay, gao2019effect, narayanan2010temperature, zhu2015strong},   because multipolar orbital order and/or complex multi-orbital arrangements  favor charge localization. A representative material of such Mott insulators   is  the  5$d^1$ double perovskite Ba$_2$NaOsO$_6${} \cite{stitzer2002crystal, steele2011low}  that evolves  to  the 5$d^2$ %(Ba$_2$CaOsO$_6$) 
 configuration upon charge doping. 

%Charge doping transforms it from  the  5$d^1$ to  the 5$d^2$ (Ba$_2$CaOsO$_6$) configuration. 

 %Ba$_2$NaOsO$_6$ is a  Os$^{7+}$ Mott insulator that displays a seemingly contradictory combination of a weak ferromagnetic moment ($\sim$0.2  $\mu_B$ / 
 This  5$d^1$  Os$^{7+}$ Mott insulator displays a seemingly contradictory combination of a weak ferromagnetic moment ($\sim$0.2  $\mu_B$ / 
 formula unit) below \mbox{$T_{C} \approx 6.8\; \textrm{K}$} and a negative Weiss temperature \cite{erickson2007ferromagnetism}. Its weak moment at low temperature derives from an exotic canted-antiferromagnetic (cAFM) phase that is preceded by a broken local point symmetry (BLPS) state \cite{Lu_NatureComm_2017, Liu_Physica_2018, PhysRevB.100.245141}. %Ba$_2$CaOsO$_6$, on the other hand, is a 5$d^2$ Os$^{6+}$ double perovskite that remains in a Mott insulating state that possibly hosts  a complex ferro-octupolar order instead of a simple N\'eel AF order \cite{yamamura2006structural, thompson2014long, 
 Fully doped, Ba$_2$CaOsO$_6$, on the other hand, is a 5$d^2$ Os$^{6+}$  Mott insulator   that possibly hosts  a complex ferro-octupolar order instead of a simple N\'eel AF order \cite{yamamura2006structural, thompson2014long, maharaj2020octupolar,maharaj2020octupolar,paramekanti2020octupolar}. Interestingly, recent  theoretical work  reveals different types of multipolar orders in $5d^2$ Mott insulators, such as ferri-octupolar \cite{lovesey2020lone}, ferro-octupolar \cite{pourovskii2021ferro,voleti2021octupolar}, and antiferro-quadrupolar ordering \cite{khaliullin2021exchange,churchill2021}. 
%
%
%\begin{center}	
 %
%%%%%%%%%%%%%%%%%%%%%%%%%%%%%%%%%%%%%%
\begin{figure*}[t]
 % \vspace*{-0.3cm}
\begin{minipage}{0.98\hsize}
%%%%%%%%%%%%%%%%%%% F I G U R E 1%%%%%%%%%%%%%%%%%%%%
 \centerline{\includegraphics[scale=0.36]{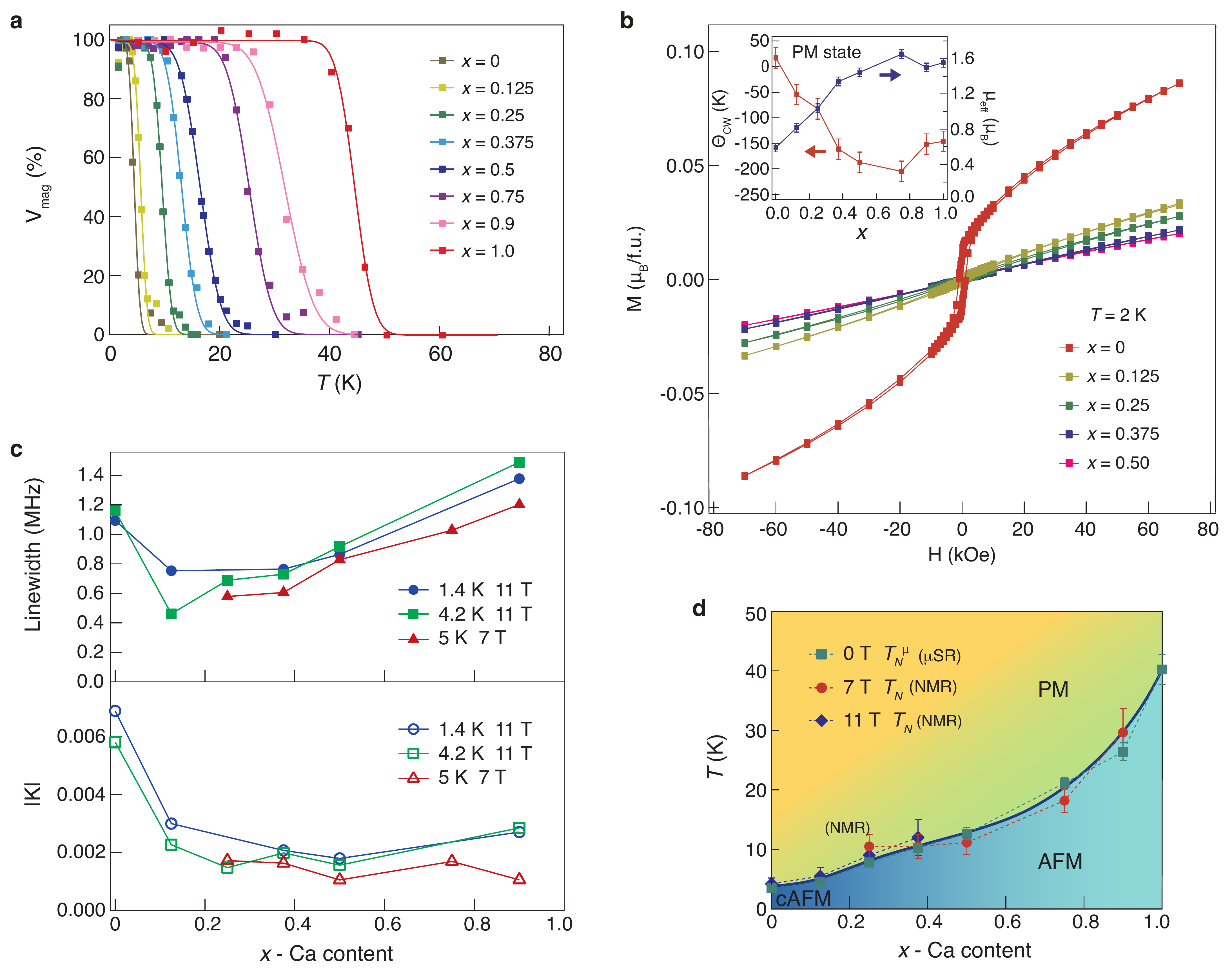}}%fig_NMRspectra_phasediag.png}} %%%%%%%%%%%%%%%
%%%%%%%%%%%%%%%%%%%%%%%%%%%%%%%%
\begin{minipage}{.99\hsize}
 \vspace*{-0.2cm}
\caption[]{\label{fig:fig1} \small %(Color online)
{\bf Magnetic state evolution as a function of charge doping ($x$) in \BNCOO.} \textbf{(a)} Magnetic volume fraction extracted from \uSR{} asymmetry for all doping levels. The magnetic transition temperature is defined as the 90\% filling of the magnetic volume  and increases monotonically with increasing Ca doping. \textbf{(b)} Magnetization as a function of applied magnetic field at \mbox{2 K}. The results of high temperature (\mbox{$T \gtrsim 50\; \textrm{K}$}) Curie Weiss fittings for magnetic susceptibility measurements in the PM state are shown in the inset. \textbf{(c)}  $^{23}$Na NMR spectral linewidth (top) and absolute value of Knight shift (bottom) as a function of doping concentration at various temperatures. \textbf{(d)} Magnetic phase diagram. %of \BNCOO. 
Markers denote magnetic transition to the canted AFM and collinear AFM state for zero-field \uSR{} and high field NMR measurements. Solid line serves as a guide to the eye. Typical error bars are on the order of a few percent and not shown for clarity in panels \textbf{a - c}.} %Temperature evolution of $^{23}$Na spectra for Ca 12.5\% at H = 11 T
 \vspace*{-0.2cm}
\label{fig1}
\end{minipage}
\end{minipage}
\end{figure*}
%%%%%%%%%%%%%%%%%%%%%%%%%%%%%%%%%%%%%%
%
%\end{center}
%
  %  
   %\vspace*{-0.2cm}
%
%
% Furthermore,  the 5$d^2$  Mott-insulators,  exhibit rich spin-orbital physics due to the complex interplay between relativistic SOC, Kugel-Khomskii type spin-orbital exchange, and Jahn-Teller orbital-lattice coupling. The interplay, and consequently the underlying physical ground state, is controlled by the multiplet structure of the constituent ions,  the nature of the chemical bonds in the crystal, and its symmetry. This complexity often leads to intricate  quantum ``hidden'' orders, elusive to most standard experimental probes. 
%
Microscopic study of the magnetic and structural  properties of the double-perovskite 5$d$ compounds with cubic symmetry, as presented here,   provides essential guidance for the development of the relevant theoretical framework for the description of   Mott insulators with strong SOC.  Once identified, such a theoretical framework can  be extended to more intricate lattices, such as the honeycomb and/or triangular lattice, where novel types of exotic quantum orders can be stabilized \cite{jackeli09mott,PhysRevResearch.4.013062,khaliullin2021exchange}.

 %%%%%%%%%%%%%%%%%%%%%%%%%%%%%%%%%%%%%%
\begin{figure*}[t]
 % \vspace*{-0.3cm}
\begin{minipage}{0.98\hsize}
%%%%%%%%%%%%%%%%%%% F I G U R E 2 %%%%%%%%%%%%%%%%%%%%
 \centerline{\includegraphics[scale=0.28]{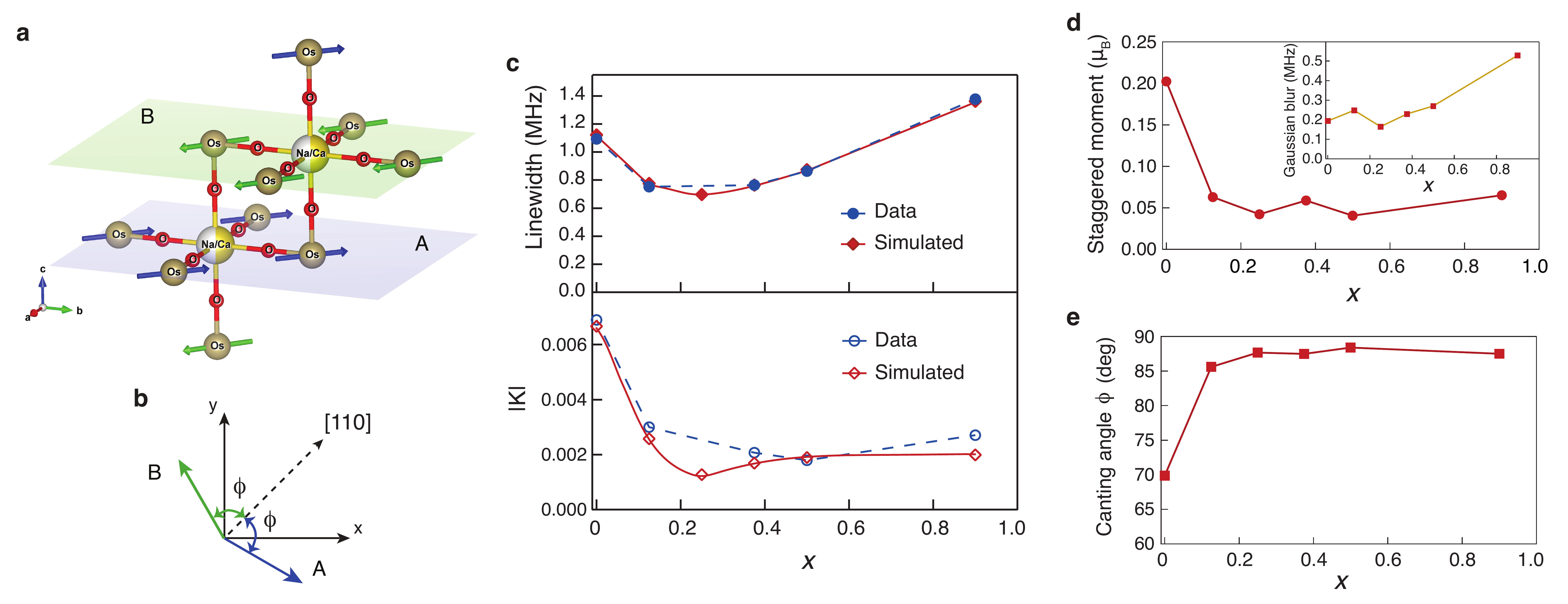}} %%%%%%%%%%%%%%%
%%%%%%%%%%%%%%%%%%%%%%%%%%%%%%%%
\begin{minipage}{.99\hsize}
 \vspace*{-0.2cm}
\caption[]{\label{fig:fig2} \small %(Color online)
{\bf Doping evolution of the staggered moment in the low temperature magnetic  state.}
\textbf{(a)} Schematic of the spin model used to fit the NMR observables. Different colors of the arrows denote different spin environments at the Os sites. The two planes with distinctly oriented moments from sub-lattice A and B are shown in different shades. \textbf{(b)} Schematic of the canted spin arrangement by angle $\phi$ with respect to the [110] direction  in the XY plane. \textbf{(c)} Simulated and measured NMR spectra linewidth and Knight shift at $T =1.4$ K and $H=11$ T as a function of Ca doping $x$ in \BNCOO. \textbf{(d)} Simulated evolution of the staggered moment, defined as the projection of moments  from two sub-lattices, A and B, along the applied field, as a function of doping in the magnetic state. The Gaussian blur, used to properly account  for  magnetic broadening, of simulated spectra is shown in the inset. 
The  blur    increases abruptly for $x=0.9$. This might be related to the increased inhomogeneity of the local magnetic field environment at the Na nuclei site. Details of this simulation can be found in Supplementary Note 6.   
\textbf{(e)} Simulated evolution of the canted angle, defined as  the angle between the sub-lattice FM spin orientation and the [110] easy axis, as a function of doping in the magnetically ordered  state. Typical error bars are on the order of a few per cent and not shown for clarity. Solid and dashed line serves as guide to the eye. }
  \vspace*{-0.2cm}
\end{minipage}
\end{minipage}
\end{figure*}
%%%%%%%%%%%%%%%%%%%%%%%%%%%%%%%%%%%%%%
%
%\end{center}

       Here, we present a comprehensive study of the effect of charge doping on a Mott insulator with both strong electron correlations and  SOC,  represented by the double perovskite Ba$_2$NaOsO$_6$, from $5d^1 \rightarrow 5d^{2}$.  
Specifically, we investigate the magnetic field-temperature ($H{-}T$) phase diagram  evolution as a function of charge doping $(x)$ achieved by   Na$^+$/Ca$^{++}$ heterovalent substitution in \BNCOO,  %{} through the Na$^+$/Ca$^{++}$ partial substitution, 
 employing magnetic resonance techniques.   We find that the system remains insulating at all doping levels, implying that the dopants form an inhomogeneous electronic state.
 We compiled a magnetic and structural phase diagram for dopant concentrations ranging from $x = 0 \rightarrow 1$. %$x = 0 \rightarrow 0.9$. 
 The  insulating magnetic ground state evolves from canted antiferromagnetic (AF) \cite{Lu_NatureComm_2017} to N{\'e}el nearly-collinear AF state (hereafter referred to as collinear AF state for brevity) for dopant levels  exceeding $\approx 10 \%$.  
Analyzing   the complex broadening of the $^{23}$Na NMR spectra,  which onsets well above the magnetic transition, and temperature dependence of NMR shift,  
we establish that a cubic to orthorhombic local distortion of the O-octahedra is present  for all   compositions \cite{PhysRevB.97.224103,cong2020first}. 
The local distortion is the  signature of a  BLPS phase, identified  as  a precursor to the magnetic state  in the single crystals of  \BNOO{} \cite{Lu_NatureComm_2017}, and not a trivial consequence of a simple structural phase transition.   
The observation of the   breaking of the local cubic symmetry and concurrent development of the NMR shift anisotropy for the entire range of dopings investigated  implies  that this symmetry breaking    is driven by a multipolar order, most-likely of the antiferro-quadrupolar type \cite{khaliullin2021exchange,churchill2021}. Remarkably, we find that 
  the cubic to orthorhombic local distortion occurs independently of the exact nature of the low temperature magnetic state, signaling that the presence of canted moments is not the  sole consequence of the BLPS \cite{Mosca21}. 
 In summary, our findings evidence that local distortions persist in the doped samples and that they favor the onset of an  antiferro-quadrupolar order. 
 
% Our findings    elucidate % the delicate interplay between electron correlations, spin-orbit entanglement, and charge and orbital degrees of freedom  on 
%   the intricate exchange interactions between effective pseudospins carrying higher-order multipoles which dictate the nature of the 
%exotic magnetic ground states \cite{doi:10.7566/JPSJ.90.062001,witczak2014correlated}.  

 \vspace*{0.2cm}
\noindent {\bf Results}

 We now describe details of our  systematic study of \BNCOO{} through the partial heterovalent substitution of monovalent Na with divalent Ca for \mbox{$0 \leq x < 1$}, performed  to  better understand the effects of doping    and to elucidate  the competing interactions that drive distinct magnetic ground states % in the Os based Mott insulators 
 %as compound evolves from the $5d^1 \rightarrow 5d^{2}$ configuration. 
  utilizing muon-spin relaxation ($\mu $SR), nuclear magnetic resonance (NMR), and magnetization measurements. % are used to probe a set of   powder compounds synthesized at different doping concentrations. 
We found that the insulting state persists at all doping concentrations despite the injection of electrons and an evolution into the AFM state. This finding  is based on thorough examination of the   response of the NMR resonant circuit.

\noindent \textbf{Magnetic state - $\mu$SR magnetic volume fraction.} First, we consider the evolution of the magnetic ground state of \BNCOO{} as a function of the Na/Ca substitution $(0 \leqslant     x \leqslant 1)$  %from the $5d^1$ $(x=0)$ to the $5d^2$ $(x=1)$ configuration 
 as probed by  zero field muon spin relaxation (ZF-$\mu$SR) measurements. 
In the absence of an external field (\mbox{$H=0\; \textrm{T}$}), the spin $I=1/2$ muon implanted in the sample precesses around the spontaneous local magnetic field arising from the magnetically ordered state at the muon site. The muon precessions are reflected in damped oscillations   of the muon asymmetry decay, probing the fraction of precessing muons, which in turn is proportional to the magnetic volume (see methods). Our ZF-muSR asymmetry measurements for the end members 
$x=0$   and $1$   are in agreement with those previously  reported in \mbox{Refs. \cite{steele2011low,lovesey2020lone}}.
In \mbox{Fig. \ref{fig:fig1}a},  we plot the temperature evolution of the magnetic volume fraction $V_{mag}$ as a function of doping. 
 We find that   samples of all concentrations display a magnetic transition as the volume fraction approaches 100\% in the low temperature limit. 
 The  transition temperature into a magnetically ordered state, $T_N^\mu$, defined to be at $V_{mag} = 90\%$,  grows monotonically from approximately  
 \mbox{$T= 5\; \textrm{K}$} to \mbox{40 K} as increasing doping    induces a configuration change  from the $5d^1$     to the $5d^2$ \cite{kesavan2020doping}, as illustrated in  \mbox{Fig. \ref{fig:fig1}d}. %Moreover, the relative  width of the  temperature range over which the $V_{mag}$ rises from zero to 100\% increases as a function of doping. 
% This implies that the width of the transition region into the long range ordered (LRO) magnetic state, \ie the inhomogeneity associated with the $T_N$,  grows as the %system evolves towards $5d^{2}$. 

% 

% \begin{center}	
 %
%%%%%%%%%%%%%%%%%%%%%%%%%%%%%%%%%%%%%%
\begin{figure*}[t]
 % \vspace*{-0.3cm}
\begin{minipage}{0.98\hsize}
%%%%%%%%%%%%%%%%%%% F I G U R E 3 %%%%%%%%%%%%%%%%%%%%
 \centerline{\includegraphics[scale=0.35]{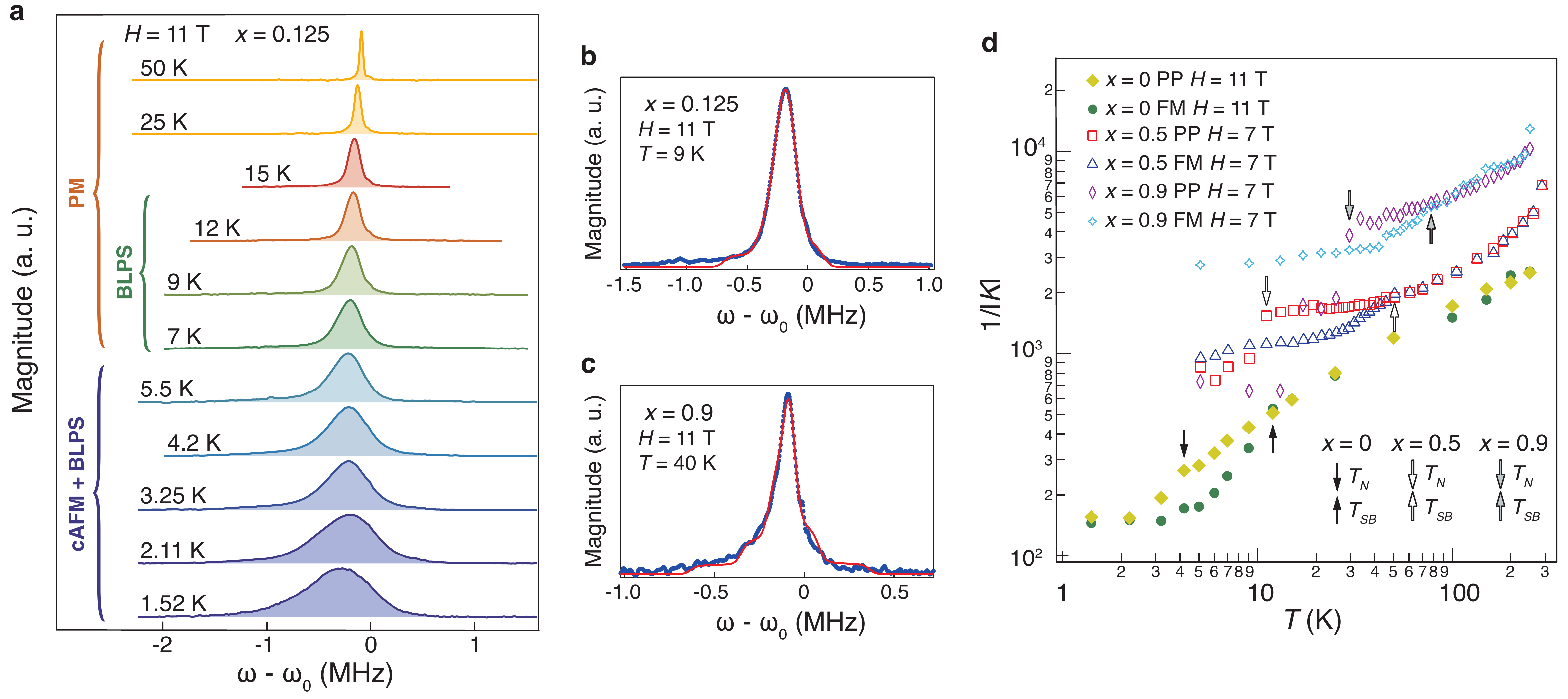}} %%%%%%%%%%%%%%%
%%%%%%%%%%%%%%%%%%%%%%%%%%%%%%%%
\begin{minipage}{.99\hsize}
 \vspace*{-0.2cm}
\caption[]{\label{fig:fig3} \small %(Color online)
{\bf NMR spectral evidence of broken local point symmetry.}
\textbf{(a)} Temperature evolution of $^{23}$Na spectra for x=0.125 at H = 11 T. $^{23}$Na powder NMR spectrum simulation results at H = 11 T for \textbf{(b)} x = 0.125 and \textbf{(c)} x = 0.9 in the BLPS phase. \textbf{(d)} Representative NMR Knight shift as a function of temperature at 11 T and 7 T. Arrows indicate the corresponding structural $T_{SB}$ and magnetic $T_{N}$ transition temperatures. $x=0.9$ (Ca=90\%) data is  vertically offset for presentation clarity. Typical error bars are on the order of a few per cent and not shown for clarity. }
  \vspace*{-0.2cm}
\end{minipage}
\end{minipage}
\end{figure*}
%%%%%%%%%%%%%%%%%%%%%%%%%%%%%%%%%%%%%%
%
%\end{center}

\noindent \textbf{Magnetic state - magnetization.} We have also performed magnetization measurements  to get a better insight into the nature of the magnetic transitions observed through ZF-\uSR{}.  In \mbox{Fig. \ref{fig:fig1}b}, we plot %representative 
magnetization curves  as a function of an applied magnetic field for \mbox{$T=2\; \textrm{K}$}. The $x = 0$ sample  displays a non-linear field dependence with a characteristic S shape and a small hysteretic behavior consistent with a moderately weak ferromagnetic character due to the significant moment canting in the cAFM phase. This is in agreement with the magnetization behavior observed in \BNOO{} single crystals \cite{erickson2007ferromagnetism}. This hysteretic behavior is rapidly suppressed with charge doping 
and is effectively undetectable for doping exceeding    $x \sim 0.1$. Magnetic susceptibility measurements  were performed as a function of temperature at \mbox{$H= 0.1\; \textrm{T}$} for all   samples. The resulting magnetic susceptibility in the high temperature paramagnetic (PM) region fits well to a Curie-Weiss (CW) function plus a small temperature independent contribution. The resulting CW temperature ($\theta_{CW}$) and effective moment per formula ($\mu_{eff}$) are displayed in the inset to \mbox{Fig. \ref{fig:fig1}b}. The values of  the end members are in very good agreement with those previously reported for \BNCOO{} at $x=0$ $(\mu_{\rm {eff}} = 0.6\mu_B)$ and $x=1$ $(\mu_{\rm {eff}} = 1.6\mu_B)  $\cite{stitzer2002crystal,erickson2007ferromagnetism,yamamura2006structural,thompson2014long}.
Furthermore, the extracted effective moments increase smoothly as the system evolves from the $5d^1$ to $5d^2$ configuration, while $\theta_{CW}$ becomes more negative. 
The extracted effective moment for both configurations is significantly suppressed from the theoretical value expected if SOC were negligible ($\mu_{\rm {eff}} = 1.73\,\mu_B$ and $\mu_{\rm {eff}} = 2.83\,\mu_B$ for $5d^1$ to $5d^2$ configurations, respectively) and is closer to the expected moments in the  infinite SOC limit ($\mu_{\rm {eff}} = 0 \mu_B$ and $\mu_{\rm {eff}} = 1.25 \mu_B$ for $5d^1$ and $5d^2$ configurations, respectively) \cite{Chen_PRB_2010,PhysRevLett.118.217202,Chen:2011}. 
The experimentally determined values of   $\mu_{\rm {eff}}$ being   significantly reduced from those expected for negligible SOC  limit indicates the presence of strong SOC (in agreement with predicted SOC coupling lambda $\sim 0.3\, \rm{eV}$), while 
their  being larger than those in the limit of infinite SOC can be attributed  to the hybridization of Os $d$ and oxygen $p$ orbitals, with extra moments coming from the $p$ orbitals, in agreement with predictions of \mbox{Ref. \cite{PhysRevB.106.035127}}.  
The most likely origin of the observed  effective moment increase with doping $(5d^1 \rightarrow 5d^2)$ is the enhancement of the spin quantum number. 
This is because  the spin $(S)$ is predicted to increase from $S=1/2$  to $S = 1$ as doping changes from $x=0 \rightarrow   x=1$   \cite{PhysRevB.106.035127}.

\noindent \textbf{Magnetic state - NMR.}  In order to obtain insight into the microscopic nature of the magnetically ordered  state throughout the doping evolution, we
 performed detailed analysis of the $^{23}$Na NMR spectra (see Methods).  
That is, we analyzed  how the spectral shape   changes 
 in the magnetically ordered state, \ie measured below  $T_N$, as a function of the Ca content $x$. 
 In  \mbox{Fig. \ref{fig:fig1}c}, we plot the   absolute value of the Knight shift,  deduced from  first moment of the NMR spectra, 
  and the linewidth of the spectra as a function of $x$.  
 The magnitude of the  Knight shift  rapidly decreased on introduction of  Ca dopants ($x > 0$) and remains  nearly constant for higher doping ($x \geqslant 0.1$).  
 The absolute value of  the    shift  is a measure of the   local   magnetic moment projected along the external magnetic field direction. In the magnetically ordered canted state, the shift is proportional to the projection of the non-compensated magnetic moment along the applied magnetic field, as was demonstrated by  
 $^{23}$Na NMR measurements on \BNOO\ single crystals \cite{Lu_NatureComm_2017, Liu_Physica_2018}.  
 Therefore, the observed abrupt decrease of the shift upon the introduction of dopants indicates that the addition of charge effectively quenches the  FM component,  
  while further doping ($x \geqslant 0.1$) leads to a more uniform distribution of the projected moments. 
   Furthermore  upon the  introduction of dopants ($x > 0$),  the linewidth,  which  reflects the distribution of the magnetic fields projected along the  %external 
    applied magnetic field, exhibits the 
   same abrupt decrease as the NMR shift,  in agreement with the suppression of the canted nature of the magnetic order upon charge doping.  
       In the collinear AFM state,  the spectral linewidth is qualitatively proportional to the size of the local  ordered magnetic moment. Thus, the smooth increase of the linewidth observed for  $x > 0.1$ in \mbox{Fig. \ref{fig:fig1}c}, indicates a progressive rise of the ordered magnetic moment as the charge concentration approaches $x = 1$, \ie  the $5d^2$ configuration 
 (full quantitative analysis of the linewidth is presented in the next section).  We point out that   one would naively expect that the linewidth  monotonically increase with doping since it introduces inhomogeneity in the crystal, which is not what was observed here because the NMR linewidth reflects intrinsic inhomogeneities of the magnetic ground state. 
 Both the magnetization and the low temperature NMR measurements are consistent with a picture where the canted magnetic state rapidly evolves into  
% Consistent with the previously described results from magnetization measurements,      
%  the doping dependence of the NMR shift and the linewidth in the low temperature magnetic phase  reveals that the canted magnetic state rapidly evolves into 
  the collinear AFM phase  upon the introduction of charge and that the ordered magnetic moment increases  as a function of Ca doping.

\noindent \textbf{Microscopic nature of the magnetic state.} To investigate the direct effect of doping on the nature of the staggered magnetic moments, \ie a two sublattice canted antiferromagnetic order observed in  \BNOO{} single crystal \cite{Lu_NatureComm_2017}, we have simulated the powder spectrum corresponding to the  two sublattice cAF order identified in  \mbox{Ref. \cite{Lu_NatureComm_2017}} (Methods). The input parameters for the simulations include the    electronic spin moment ($\vec{S}$) and the  hyperfine coupling tensor 
 ($\mathbb{A}$). For each Ca doping concentration,  we fix the value of  $\vec{S}$  to be that of the effective moment  ($\vec{S} = \mu_{{\rm eff}}$) deduced  from susceptibility measurements (the inset to \mbox{Fig. \ref{fig:fig1}b}).  
  Simulated spectra are then fitted to the observed ones with canting angle, and consequently staggered moment, as a fitting parameter. 
  Here,  the canting angle is that between the sub-lattice FM spin orientation and the [110] easy axis, while the staggered moment refers to the projection of the moments
    from the two sub-lattices, A and B, along the applied field direction. 
  The simulation results  are summarized in \mbox{Fig. \ref{fig:fig2}}. 
  We find that both, the canting  angle and the staggered moment, change abruptly  upon charge doping for $x > 0.1$. For $x=0$,  we  obtain the best fit for 
a canting  angle of $68\degree$,   consistent with that reported for pure \BNOO{} single crystal  \cite{Lu_NatureComm_2017}.
For $x > 0.1$, the canting angle rapidly approaches  $90\degree$, the value associated with a collinear AFM state (see \mbox{Fig. \ref{fig:fig2}d}). 
The deduced  canting   angles and staggered moments  serve as input parameters to calculate  the doping evolution of the   Knight shift and linewidth associated with the  local spin arrangement depicted in \mbox{Fig. \ref{fig:fig2}a}.  The calculated evolution of the shift and the linewidth is in an excellent agreement with   observations, as shown in \mbox{Fig. \ref{fig:fig2}c}.
As $x$ increases, the Knight shift decreases as a direct consequence of the increase of the canting   angle, as depicted in  \mbox{Fig. \ref{fig:fig2}b}.
On the other hand,   the linewidth decrease with increasing $x$ is associated with  the weakening of the off-diagonal components of the hyperfine coupling tensor,    for the canting  angle   $\sim 90\degree$.  Furthermore   in the inset to \mbox{Fig. \ref{fig:fig2}d} 
  we plot the Gaussian blur, which is used to account for the inhomogeneous magnetic broadening of the measured spectra,  and its abrupt   increase with doping  approaching $x \rightarrow 1$ indicates that injected charge  is inhomogeneously distributed.

\noindent \textbf{Broken local point symmetry (BLPS).} Next, we investigate the effect of  charge doping on the   BLPS phase  \cite{Lu_NatureComm_2017}. 
The onset  temperature of the visible  broken local point symmetry is determined from the NMR Knight shift.  
We   then proceed to analyze NMR powder spectral shapes at temperatures above  the transition to the magnetically ordered state  to determine the nature of the local crystal symmetry. 

This subtle symmetry breaking has eluded previous diffraction measurements \cite{erickson2007ferromagnetism}, but is well reflected in the distortion of the NMR spectra by the unbalanced  spectral weight distributed towards a lower frequency with respect to its main peak associated  with the cubic symmetry.  A direct phenomenological way to detect and estimate the distortion of the spectra is by comparing the temperature evolution of the frequency of the peak position of the spectra to the first moment of the frequency distributions. In \mbox{Fig. \ref{fig:fig3}d}, we display representative data sets  for the comparison of the Knight shift 
obtained relative to the frequency $\omega^i$ of the first moment ($i=fm$) and peak   ($i=pp$) of our NMR spectral lines (see Methods). The clear bifurcation in 1/$|K|$ between the shift obtained by using the peak position and first moment marks the structural phase transition to the BLPS phase, $T_{SB}$. Upon further cooling, a sharp drop of 1/$|K|$ is   observed, denoting the transition temperature  $T_N$ into the  magnetic state (see Methods). The magnetic transition temperature, $T_N$,  obtained from the  1/$|K^{pp}|$  drop, is consistent with that determined from  the peak of the  NMR relaxation rate (not shown) and   from  ZF-muSR measurements (\mbox{Fig. \ref{fig:fig1}d}). These observations  indicate that below $T_N$,  an AF order is  formed. 
In an AF ordered state,  net projected moment along the applied field is significantly reduced, and vanishes in the case of collinear AF order,   resulting in   the observed drop in the Knight shift.   

%\begin{center}	
 %
%%%%%%%%%%%%%%%%%%%%%%%%%%%%%%%%%%%%%%
\begin{figure}[t]
 % \vspace*{-0.3cm}
\begin{minipage}{0.98\hsize}
%%%%%%%%%%%%%%%%%%% F I G U R E 4 %%%%%%%%%%%%%%%%%%%%
 \centerline{\includegraphics[scale=0.35]{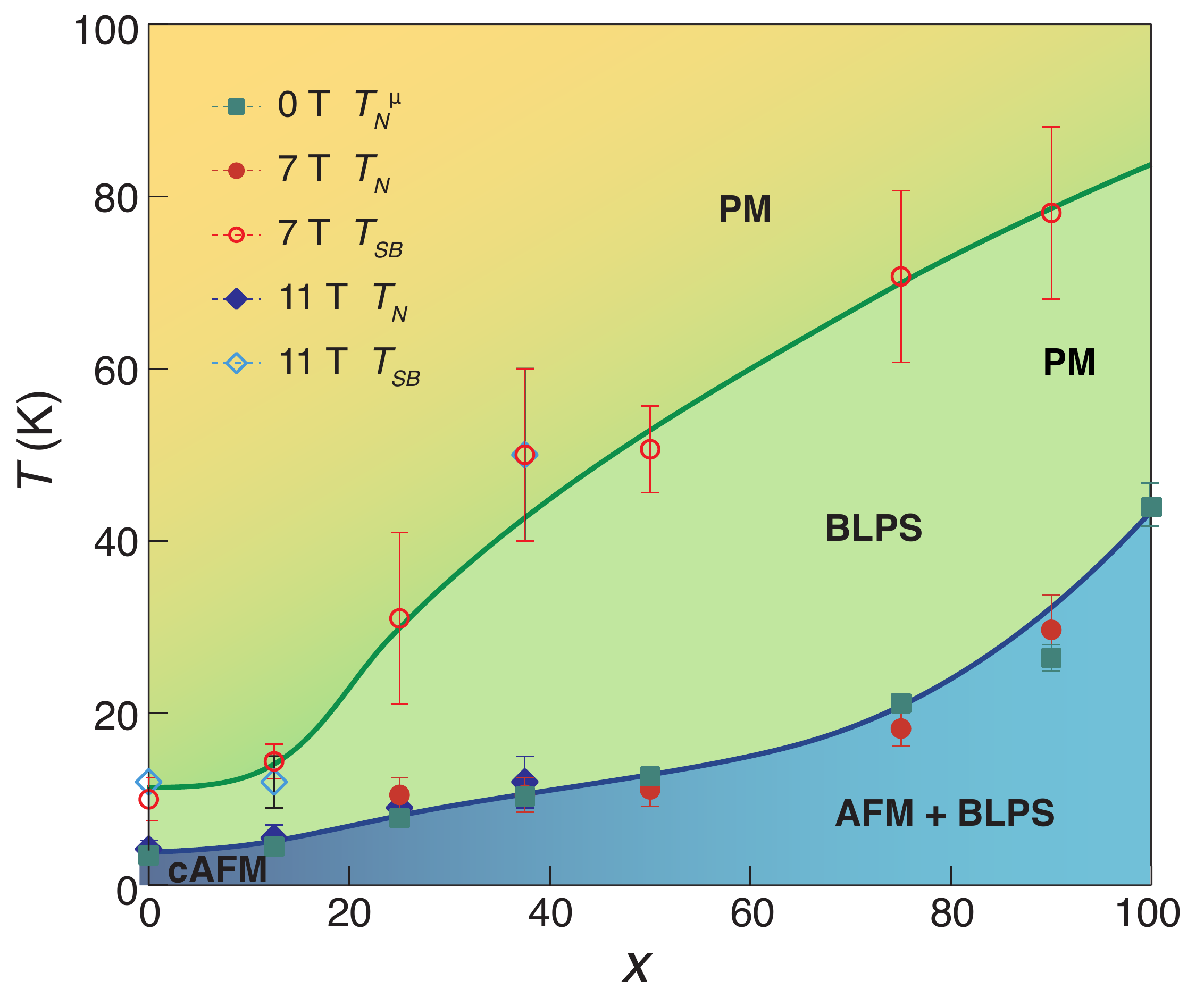}}%fig_NMRspectra_phasediag.png}} %%%%%%%%%%%%%%%
%%%%%%%%%%%%%%%%%%%%%%%%%%%%%%%%
\begin{minipage}{.99\hsize}
 \vspace*{-0.2cm}
\caption[]{\label{fig:fig4} \small %(Color online)
\textbf{Phase diagram of \BNCOO.} Solid markers denote the magnetic transition to the canted AFM and collinear AFM state for zero-field \uSR{} and high field NMR measurements. Open markers denote structural transitions into the BLPS phase. Solid lines serve as a guide to the eye.}
 \vspace*{-0.2cm}
\end{minipage}
\end{minipage}
\end{figure}
%%%%%%%%%%%%%%%%%%%%%%%%%%%%%%%%%%%%%%
%
%\end{center}

\noindent \textbf{Nature of the broken symmetry.} 
In  \mbox{Fig. \ref{fig:fig3}a}, we plot 
  $^{23}$Na NMR spectra of \BNCOO{}   for \mbox{$x = 0.125$} at \mbox{$11\; \textrm{T}$}. The cubic paramagnetic (PM) state is characterized by the narrow symmetric    spectra, as expected in the highly symmetric cubic PM phase. The  BLPS arises in the intermediate temperature range  and is marked by  asymmetric spectra with a more pronounced tail at   lower frequencies. At low temperatures, in the magnetically ordered phase,  the BLPS phase 
  coexists with magnetism  revealed by asymmetry of the NMR spectra.  As elaborated  in the Supplementary Note 4, 
  a cubic local environment at the nuclear site must lead to a symmetric spectrum, while only non-cubic  local  symmetries, such as  tetragonal or orthorhombic, can generate asymmetric lineshapes.  We have performed detailed simulations of the $^{23}$Na NMR  powder pattern  spectra in the presence of  a quadrupolar interaction with the electric field gradient (EFG) and an anisotropic Knight shift $\mathbb{K}$ (see Methods \& Supplementary Note 4), following the notation of \cite{Vachon08}. We find that above $T_{N}$, in the PM \& BLPS phases,    the resulting powder NMR spectra must reflect the symmetry of the $\mathbb{K}$ and EFG tensors. Therefore, the shape of the NMR spectra provides  precise information about any deviation from the cubic symmetry.
Indeed,  our systematic analysis   of the measured spectra demonstrates  that the best fits can only be  achieved by  using orthorhombic distortions,   in agreement with findings  in  \BNOO\ single crystals \cite{Lu_NatureComm_2017}.  

In \mbox{Fig. \ref{fig:fig3}b \& 3c} we illustrate   representative  results of our  $^{23}$Na NMR powder  spectra simulations in the BLPS phase (details of the simulation are given in Supplementary Note 4).  By  fitting the powder spectra, we deduce that  the EFG and $\mathbb{K}$ tensors are orthorhombic and collinear to one another. 
  These findings reveal that the main signature of the  BLPS phase is not only  the cubic symmetry breaking, but also the concurrent  development of a collinear 
  $\mathbb{K}$ anisotropy,  
 indicating that the BLPS is a consequence of a multipolar order formation.  That is, the deduced $\mathbb{K}$ anisotropy implies that the BLPS does not consist of a simple structural distortion but rather involves distortions of magnetic super-exchange paths, plausibly induced by a formation of a multipolar ordering.  Therefore, the BLPS is most likely an anisotropic multipolar phase.  

\noindent \textbf{Phase diagram.} Both $T_{SB}$ and $T_N$ obtained from $^{23}$Na NMR at 7 T and 11 T are displayed in the temperature versus doping phase diagram plotted in \mbox{Fig. \ref{fig:fig4}}.  The results show that the BLPS phase, a precursor to the magnetic state, is an intrinsic characteristic of these Mott-insulating metal oxides and persists to higher  temperature, up to about 80 K, when approaching the $5d^2$ configuration. This orthorhombic local point  distortion of the octahedra is a clear signature   that these materials are intrinsically dominated by low temperature anisotropic spin-lattice interactions, an essential ingredient to be included in any microscopic quantum theory of Dirac-Mott insulators.

 % \section{Discussion}
%\label{Disc}
%  \vspace*{-0.20cm}
  
% { \it Discussions.}
% \label{Disc}

\vspace*{0.2cm}
 \noindent {\bf Discussions}

%\vspace*{0.2cm}
%\noindent {\bf Conclusion}

We have performed a microscopic investigation of the transformation   of  a $5d^{1}$ double perovskite Mott insulator into a $5d^{2}$ configuration  by    charge doping.     We observed that the system  remains insulating while %the  size of the temperature range over which the magnetic volume fraction rises to 100\%   and 
 the NMR linewidth  (Gaussian blur)  increase as  doping approaches $x \rightarrow 1$.  % when  $5d^{1} \rightarrow 5d^{2}$  as a function of doping. 
These  findings indicate that  the injected charges are not uniformly distributed   into   the $5d^{1}$ double perovskite, rather they are inhomogeneously trapped, most likely on the Os sites that convert the system to $5d^{2}$ configuration. The formation of polarons, quasiparticles formed by the coupling of excess charge with ionic vibrations, has been    
recently proposed as another  plausible mechanism for dopant trapping \cite{2022polaron,PolaronsCF21}.

Our magnetization measurements reveal that  % the system evolves towards the infinite SOC limit and 
 AF exchange interactions become enhanced as charge doping alters its configuration from $5d^{1} \rightarrow 5d^{2}$.   Furthermore, detailed analysis of the     NMR shift and lineshapes in the magnetically ordered state  reveals that magnetism coexists with the  BLPS  phase, and that it evolves from a canted AFM state to a collinear AF for dopant levels  exceeding   $(x > 0.1)$. In $5d^{1}$ \BNOO,  a cAF order  was identified to arise from the BLPS, as a result of the interplay of  electron correlations and the degree of Jahn-Teller distortions \cite{Mosca21}.  That is, the collinear AF order cannot be stabilized in the absence of cubic symmetry in  the $5d^{1}$ compound. 
%\ie the canting angle was determined by 
Our finding that a collinear AF order coexists with the BLPS for $x>0.1$ suggests that the theoretical results of \cite{Mosca21} need to be extended to include effects of enhanced electron correlations, SOC,  and Jahn-Teller  interaction  with doping. Therefore, our result 
show that  charge doping profoundly alters the interplay between spin-exchange interactions, Jahn-Teller distortions, and electronic correlations as  the system evolves away from   the $5d^{1}$ configuration, and provides pertinent constraints to guide   the  development of microscopic models of Mott insulators with SOC.  

 Through a comprehensive  analysis of the   temperature dependence of the NMR shift and lineshapes, we established that a BLPS  phase
  occupies an increasing  portion of the  ($H{-}T$) phase diagram with increased doping. 
  This finding is seemingly in contradiction with those  in  \cite{maharaj2020octupolar,CMThompsonetal} reporting a single transition into N{\'e}el order % with a small moment with an upper limit of $0.06 \, \mu_{{\rm B}}$    
  at $T^{*}   \approx 50$ K, and no evidence of deviations from cubic symmetry  in \BCOO. It was proposed that these rather unusual results below $T^{*}$   in \cite{maharaj2020octupolar}  may be  reconciled by the emergence of time-reversal symmetry breaking, ferro-octupolar order \cite{pourovskii2021ferro,paramekanti2020octupolar,voleti2021octupolar}. 
 Although  the diffraction data in  \cite{maharaj2020octupolar} is high intensity, it is not   obvious that  its resolution was sufficient  to detect a small deviations from cubic symmetry like those reported here and/or seen in Ba$_2$MgReO$_6$ \cite{PhysRevResearch.2.022063}.  Improved resolution in  \cite{CMThompsonetal}   might still not be good enough to detect a deviation from cubic symmetry.  Moreover,  it is possible that the local deviations from cubic symmetry in the BLPS, as reported here, are not sufficiently coherent to drive a long range distortion observable in scattering experiments. Indeed, an antiferro order of active quadrupoles  within the $e_g$ doublet  was identified in \cite{khaliullin2021exchange} as a competing phase. 
  If the   distortions inherent to the BLPS phase are present, then the quadrupolar AFM phase is more stable. This is in contrast with the case when the  system preserves  cubic symmetry and acquires  a octupolar-FM order.  Since these two multipolar phases are very close in energy, it could be that small perturbation induce   tiny local distortions and  promote   the onset of  the quadrupolar AFM.

   Our microscopic data clearly shows that in the limit of  $x \rightarrow 1$,  the system deviates from cubic symmetry below \mbox{$T_{SB} \sim 80$ K}, while a N\'eel AF order develops below     $T_{N} \sim 30$ K, with a staggered moment of $0.05 \, \mu_{{\rm B}}$. 
  Therefore, our observations of the broken local cubic symmetry  indicate  that the BLPS phase  is driven by a formation of a multipolar order, most-likely of the antiferro-quadrupolar type \cite{khaliullin2021exchange,churchill2021}, because  ferro-octupolar order preserves cubic symmetry.   The conclusion that the BLPS phase   is of antiferro-quadrupolar type is supported by two additional findings. 
 Firstly,  the BLPS phase is characterized by an  anisotropic NMR shift tensor in addition to  orthorhombic local distortions. Secondly, the alluded inhomogeneous  nature of the charge doping  promotes quadrupolar four-spin exchange interactions, making quadrupolar phase more stable, and implying that the onset of the BLPS-quadrupolar phase should occur at a higher temperature as doping increases, consistent with presented  observations. 
   Our NMR measurements indicate that local distortions, induced by the antiferro-quadrupolar order, are inhomogeneous, \ie consists of areas of distortions of different magnitude. Such local distortions,   seen by NMR, do not coherently order and give rise to diffraction peaks here in the way they do in the related $5d^1$ compound, Ba$_2$MgReO$_6$ \cite{PhysRevResearch.2.022063}.

  One could argue that  ferro-octupolar order is not observed in our doped samples  simply because inhomogeneous dopants can induce local strain, consequently leading to  a  breaking of the local  cubic symmetry.   However,  local strain is predicted to suppress   the  octupolar ordering temperature %, $T_{SB}$,   
   because it    induces a transverse field  in  the octupolar ordering direction which promotes quantum fluctuations  \cite{voleti2021octupolar}, and thus is in contradiction with   phase diagram presented in \mbox{Fig. \ref{fig:fig4}}. 
  Therefore, our NMR findings support emergence of an antiferro-quadrupolar order in the doped $5d^{1} \rightarrow 5d^{2}$ double perovskites. The quadrupolar phase most-likely  arises as  lattice  distortions  amplify the quadrupolar interactions via Jahn-Teller interactions.

\vspace*{0.2cm}
\noindent {\bf Methods} 

\noindent {\bf Sample.} The powder samples of \BNCOO{} investigated here are the same as in Ref. \cite{kesavan2020doping}.
 Powder x-ray diffraction (PXRD) measurements were performed to test the quality of the samples and analysed with Rietveld refinement. The compositional evolution of the lattice parameter is shown to follow Vegard's Law, indicating a successful Na/Ca substitution \cite{kesavan2020doping}.

\noindent {\bf Muon spin resonance.} All $\mu$SR measurements were performed at the General Purpose Surface-Muon Instrument at the Paul Scherrer Institute in Switzerland. In  $\mu$SR measurements, spin-polarized muons implant in the powder samples and precess around the local magnetic field with a frequency given by $\nu=\gamma_{\mu} \cdot $\textbar B\textbar /$2\pi$, where $\gamma_{\mu}=2\pi\cdot135.5$ MHz/T. The muons decay with a characteristic lifetime of 2.2 $\mu$s, emitting a positron, preferentially along the direction of the muon spin. The positrons are detected and counted by a forward ($N_F (t)$) and backward detector ($N_B (t)$), as a function of time. The asymmetry function A(t) is given by
\begin{equation}
A(t)=\frac{N_B (t) - \alpha N_F (t)}{N_B (t)+\alpha N_F (t)},
\end{equation}
where $\alpha$ is a parameter determined experimentally from the geometry and efficiency of the $\mu $SR detectors. $A(t)$ is proportional to the muon spin polarization, and thus reveals information about the local magnetic field sensed by the muons.
 A typical zero-field \uSR{} spectra below and above the magnetic transition temperature for \BNCOO{} are presented in  the Supplementary Material. The \uSR{} spectra for all doping concentrations display damped oscillations at low temperatures (Supplementary Material 1), marking a transition to a state of long-range magnetic order. The spectra for  the end elements, $x = 0$ and $x = 1$, are in agreement with those previously reported \cite{steele2011low, thompson2014long}. Each individual spectrum was fitted to a sum of precessing and relaxing asymmetries given by
\begin{eqnarray}
\label{asymm}
A(t)&= \left [ \sum\limits_{i=1}^{2} A_i e^{-\frac{\sigma_i^2 t^2}{2}}cos(2\pi \nu_i t) + A_3 e^{-\frac{\sigma_3^2 t^2}{2}} \right ] \\ \nonumber &+ A_\ell e^{-\frac{t}{T_1}}.
\end{eqnarray}
The terms inside the brackets reflect the perpendicular component of the internal local field probed by the spin-polarized muons, the first term corresponds to the damped oscillatory muon precession about the local internal fields at frequencies $\nu_i$, while the second reflects a more incoherent precession with a local field distribution given by $\sigma$. The term outside the brackets reflects the longitudinal component characterized by the relaxation rate $T_1$.

The development of a magnetic phase can be probed by measuring the volume of magnetic and non-magnetic regions within our sample. This magnetic volume fraction (V$_{mag}$) can be obtained from the longitudinal (A$_\ell$) and total (A$_{tot}$) component of the polarized muons, given by the expression
\begin{equation}
\label{magvol}
V_{mag}=\frac{3}{2}\left(1-\frac{A_\ell}{A_{tot}}\right).
\end{equation}

\noindent {\bf Magnetization.} Bulk magnetization measurements were performed using a superconducting quantum interference device (SQUID) magnetometer. Isothermal magnetization measurements as a function of applied field were performed at 2 K from -70 to 70 kOe. Zero-field and field-cooled magnetic susceptibility measurements were performed from 2 K to 400 K under an applied field of 1000 Oe.

\noindent {\bf NMR.} NMR measurements were performed using high homogeneity superconducting magnets at Brown University, and the National High Magnetic Field Laboratory in Tallahassee, FL for magnetic fields exceeding 9 T. Temperature control was provided by $^4$He variable temperature inserts. $^{23}$Na NMR data ware collected using state-of-the-art, laboratory-made NMR spectrometers from the sum of spin-echo Fourier transforms recorded at constant frequency intervals. Pulse sequences of the form ($\pi/2 - \tau - \pi/2$) were used and none of the presented NMR observables depend  on the  duration of time interval $\tau$. 

\noindent {\bf $^{23}$Na NMR.} $^{23}$Na NMR is a  powerful local probe of both the electronic spin polarization (local magnetism), via the hyperfine coupling between electronic magnetic moments and the $^{23}$Na nuclear spin \mbox{$I=3/2$}, and the charge distribution (orbital order and lattice symmetry), via the quadrupolar interaction between the electric field gradient (EFG) and the $^{23}$Na quadrupole moment Q \cite{Lu_NatureComm_2017, PhysRevB.97.224103}. They affect both the first and second moments of the frequency distribution, \ie   the NMR spectra.

\noindent {\bf The Knight shift. }
The Knight shift 
is defined as $$K^i= (\omega^i - \omega_0)/ \omega_0,$$ 
where $\omega_0 =\; ^{23}\gamma \cdot H_0$, where \mbox{$^{23}\gamma$ = 11.2625 MHz/T}  and $H_0$ is the externally applied magnetic field, and $\omega^i$ is obtained from the first moment and/or peak position of the NMR spectral lines. 
This equation forms the general definition of the NMR shift $K$ in terms of the observed NMR frequency, $\omega^{i}  \equiv \gamma H_0 (1+K)$, where $K\equiv H_{loc}/H_0$ is the shift. Therefore,  $K$ is a  measure of the relative strength of the   component of the local magnetic field ($H_{loc}$) parallel to the applied magnetic field,    $H_0$.  
 In the more general case where the shift varies as a function of the orientation of $H_0$ (anisotropic shift), the scalar $K$ is promoted to a second-rank tensor $\mathbb{K}$ and the expression for the observed NMR frequency becomes \cite{Abragam1961}
 \begin{equation}\label{shift_tensor}
 \omega^{i} = \gamma H_0(1+ \hat{\mathbf{h}} \cdot \mathbb{K} \cdot \hat{\mathbf{h}}), 
 \end{equation}
where $\hat{\mathbf{h}} = \mathbf{H}_0/ H_0$ is a unit vector in the direction of the applied magnetic field.

When the quadrupole interaction is taken into account, the observed NMR frequency becomes, 
%\begin{equation}\label{quadrupole}
%\begin{split}
%$\omega^{i} \,  =  \, & \gamma (1 + K)H_0   \, + \, \\ &\omega\smallindice{Q} (m -1/2) \left( 3 \cos^2 \theta\smallindice{Q} -1 + \eta \sin^2 \theta\smallindice{Q}
% \cos 2 \varphi\smallindice{Q} \right)
%\end{split}
%\end{equation}
 $\omega^{i} \,  =   \gamma (1 + K)H_0   \, +    \omega\smallindice{Q} (m -1/2) \left( 3 \cos^2 \theta\smallindice{Q} -1 + \eta \sin^2 \theta\smallindice{Q}
  \cos 2 \varphi\smallindice{Q} \right)$
    up to second order in perturbation theory  \cite{Lu_NatureComm_2017, Abragam1961, PhysRevB.97.224103}.   
    The second term accounts for the quadrupole interaction  for each $m \leftrightarrow m \pm1$ transition and can be used   to deduce the quadrupole parameters  when the principal axes of the EFG tensor  coincide with those of the crystal, as in the case of \BNOO{}  \cite{Lu_NatureComm_2017}. 
    Here, $\theta\smallindice{Q}$ and $\varphi\smallindice{Q}$ are the angles between the applied field and the principal axes of  the electric field gradient (EFG)  defined so that  $|V_{ZZ}| \geq |V_{XX}| \geq |V_{YY}|$ and $eq \equiv V_{ZZ}$. The asymmetry parameter $\eta$ is set as 
\mbox{$\eta \equiv (V_{XX} -V_{YY})/V_{ZZ}$}.  
    The quadrupolar frequency equals to \mbox{$\omega\smallindice{Q}= 3 e^2 q Q/( \hbar 2 I(2I-1))$}, where $Q$ and $q$ are the nuclear  and electronic quadrupole moments.

To relate the quadrupolar interaction effect to the  observed NMR frequency,  we introduce the quadrupolar splitting tensor $\mathbb{W}$, such that $\Delta \omega = \sum_{\alpha} W_{\alpha} \hat{h}^2_{\alpha}$ in the coordinate system $O_{xyz}$, where $\mathbb{W}$ is diagonal. Here $O_{xyz}$ is defined by the crystalline axis of the cubic perovskite unit cell \cite{PhysRevB.97.224103},      $\alpha=\{x,y,z\}$, and $\hat{h}$ is the unit vector along the applied field. 

In  the presence of the anisotropic shift and   quadrupolar interactions, the full expression for the observed NMR frequency becomes
 \begin{equation}\label{shift_tensor}
 \omega^{i} = \gamma H_0(1+ \hat{\mathbf{h}} \cdot \mathbb{K} \cdot \hat{\mathbf{h}} + \mathbb W  \cdot \hat{\mathbf{h}}^2 ).
 \end{equation}

\noindent{\bf Calculation of NMR spectra below $T_{N}$.}
In the magnetically ordered phase, 
we model  the local magnetic field at a  Na site  as $H_{loc}= \hat {\bf{h}} \cdot \sum_{i}\mathbb{A}_{i}\cdot \vec{S}_{i}$, where $\hat {\bf{h}}$ is a unit vector in the applied field direction, $\mathbb{A}_{i}$ is the hyperfine coupling tensor  with the $i^{{\rm th}}$ nearest-neighbour Os atom, and $\vec{S}_{i}$ is its local spin moments. 
We note that contributions from dipolar effects are relatively small and are thus neglected. By performing a full lattice sum, we calculate the local fields projected  in the direction of the applied field at the Na sites for the entire single crystal. 
The corresponding NMR spectra  is a histogram of these local fields. 
  The powder spectrum is then obtained by rotating the $\hat {\bf{h}}$ over the solid angle and integrating the results (see Supplementary Note 4). 
The diagonal components of the hyperfine tensor are optimized based on the hyperfine coupling values obtained from the NMR Knight shift and magnetic susceptibility measurements (see Supplementary Figure 4), while the symmetry of the hyperfine tensor is assumed to be the same as that found in the  single crystal \BNOO{} \cite{Lu_NatureComm_2017}. 
  The quadrupolar broadening effects are accounted  for utilizing EFG parameters determined  from the intermediate temperature powder spectrum simulation (see Supplementary Note 4).

\noindent {\bf Transition temperature.}  The transition temperature $(T_N)$ from the PM to low-temperature magnetically ordered state was determined by examining    
the magnetic volume fraction $V_{mag}$ from zero-field $\mu$SR,  and in finite fields from the temperature dependence  of the NMR observables. 
That is, $T_N^\mu$ is  delineated as a temperature at which $V_{mag}$ exceeds 90\%. 
From NMR analysis, $T_N$ is defined as the onset temperature of a significant drop in the Knight shift  associated with the formation of the AF ordering.  Such $T_N$ is consistent with the one determined from the peak of the NMR relaxation rate, induced by critical fluctuations when approaching a transition into a long-range ordered magnetic state  as a function of temperature.  
 
  The onset temperature for the breaking of local cubic symmetry, $T_{SB}$, was identified by the  bifurcation in $1/|K| \, vs. \, T$ between the shifts defined relative to $\omega^i$,  corresponding to the peak position ($i=pp$)  and first moment  ($i=fm$).

%%%%%%%%%%%%%%%%%%%%%%%%%%%%%%%%%

%%%%%%%%%%%%%%%%%%%%%%%%%%%%%%%%%

%  \section{Acknowledgments}
%\label{Ack}

\vspace*{0.2cm}
\noindent  {\bf ACKNOWLEDGEMENTS} \\
\noindent We are thankful to  D. Fiore Mosca,  P. Santini, Brad Marston, and A. Paramekanti for enlightening discussions. This work was supported in part by U.S. National Science Foundation (NSF) grant No. DMR-1905532 (V.F.M.), the NSF Graduate Research Fellowship under Grant No. 1644760 (E.G.), NSF Materials Research Science and Engineering Center (MRSEC) Grant No. DMR-2011876 (P.M.T. and P.M.W.), and University of Bologna (S.S. and P.C.F.). 
Part of this work is based on experiments performed at the Swiss Muon Source S$\mu$S, Paul Scherrer Institute, Villigen, Switzerland.
The study at the NHMFL was supported by the National Science Foundation under Cooperative Agreement no. DMR-1644779  and the State of Florida. \\

%%%%%%%%%%%%%%%%%%%%%%%%%%%%%%%%%

 \noindent  {\bf AUTHOR CONTRIBUTIONS}\\
  E.G., R.C., P.C.F., A.T. , A. P. R. and G.A. obtained NMR data and performed the data analysis. S.S. lead the muon spin spectroscopy experiment for which E.G. and R.C. contributed equally to data acquisition and interpretations. P.M.T. and P.M.W. grew the samples and performed the magnetic susceptibility measurement. S.S., V.F.M. and C.F. ideated the project. S.S. and V.F.M. supervised the experiments, lead data analysis and interpretations and writing of the manuscript. R.C., G.A., and V.F.M. were involved in the simulations of the NMR data. C.F. performed first principle calculations and was involved in data interpretations. A.P.R. enabled all measurements above 10 T at the NHMFL. All authors have read the paper and approved it. All authors discussed the results and commented on and edited the manuscript.\\

%\noindent  {\bf Additional information}\\
 
%\noindent {\bf Supplementary Information} accompanies this paper at  \
 
%\noindent  {\bf Competing financial interests:} The authors declare no competing financial interests.  Readers are  welcome to comment on the online version of this article at www.nature.com/nature. All relevant data are available from the authors upon request. Correspondence and requests for materials should be addressed to V. F. M. (vemi@brown.edu) \& S. S. (s.sanna@unibo.it).\\
 
% \noindent  {\bf  Reprints and permission} information is available online at http://npg.nature.com/reprintsandpermissions/\\
 
%\noindent  {\bf How to cite this article:}  \\
 
%\noindent {\bf Publisher's note:} Springer Nature remains neutral with regard to jurisdictional claims in published maps and institutional affiliations.

\bibliographystyle{unsrt}%{ieeetr}
\bibliography{references.bib}

\end{document}